\documentclass{llncs}
\usepackage{amsmath}
\usepackage{comment}
\usepackage{amsmath}
\usepackage{amssymb}
\usepackage{amsfonts}
\usepackage{graphicx}
\usepackage{subfig}
\usepackage{color}
\usepackage{multicol}
\usepackage{multirow}
\usepackage[ruled,noline]{algorithm2e}
\usepackage{comment}
\usepackage{float}
\usepackage{enumitem}

\DeclareMathOperator{\E}{\mathbb{E}}

\renewcommand{\footnoterule}{%
  \kern -3pt
  \hrule width \textwidth height 0pt
  \kern 2pt
}

\usepackage{mdwmath}
\usepackage{mdwtab}
\usepackage{eqparbox}
\usepackage{color, soul}

\usepackage{graphicx}
\begin{document}

\title{Unified cross-modality  feature disentangler for unsupervised multi-domain MRI abdomen organs segmentation} 
\author{Jue Jiang\inst{1}
       %index{First Name, Last Name} 
  \and Harini Veeraraghavan \inst{1}}
      %index{First Name, Last Name}  

% If authors are from different institutes 
\institute{Department of Medical Physics, Memorial Sloan Kettering Cancer Center \email{veerarah@mskcc.org} }

\maketitle              
\begin{abstract}
\textcolor{black}{Our contribution is} a unified cross-modality feature disentagling approach for multi-domain image translation and multiple organ segmentation. Using CT as the labeled source domain, our approach learns to segment multi-modal (T1-weighted and T2-weighted) MRI having no labeled data. Our approach uses a variational auto-encoder (VAE) to disentangle the image content from style. The VAE  constrains the style feature encoding to match a universal prior (Gaussian) that is assumed to span the styles of all the source and target modalities. The extracted image style is converted into a latent style scaling code, which modulates the generator to produce multi-modality images according to the target domain code from the image content features. Finally, we introduce a joint distribution matching discriminator that combines the translated images with task-relevant segmentation probability maps to further constrain and regularize image-to-image (I2I) translations. We performed extensive comparisons to multiple state-of-the-art I2I translation and segmentation methods. Our approach resulted in the lowest average multi-domain image reconstruction error of 1.34$\pm$0.04. Our approach produced an average Dice similarity coefficient (DSC) of 0.85 for T1w and 0.90 for T2w MRI for multi-organ segmentation, which was highly comparable to a fully supervised \textcolor{black}{MRI multi-organ segmentation} network (DSC of 0.86 for T1w and 0.90 for T2w MRI).\footnote{\textcolor{blue}{This paper has been accepted by MICCAI2020}} 
\begin{comment}
that embeds the image content into features that when combined with the appropriate target domain style and domain code transforms the source into target image. We developed a new multi-domain adaptation methods. We disentangled each domain into content, style and domain code. The style code to learn to scale the higher level generator features. Moreover, we introduced a multi-domain structure discriminator to matching the joint distribution of image and their segmentation probability maps, which constrained both the image translation and segmentation. We compared with multiple single and multi-image translation as well as domain adaptation methods. The proposed method achieved comparable or better accuracy than other methods as well with supervised training.
\end{comment}
\end{abstract}
%\bumpup
%\bumpup
\section{Introduction}
Magnetic resonance imaging guided radiation therapy treatments require accurate segmentation, which is currently done by clinicians~\cite{kupelian2014}. Despite the availability of in-treatment-room-imaging, due to lack of fast segmentation methods, these treatments cannot be used to adapt treatments and precisely target tumors. Deep learning methods cannot be directly applied to MRI as large expert-labeled datasets are lacking. Therefore, we developed an  unsupervised multi-domain MRI (T1-weighted, T2-weighted) segmentation approach by using CT as a source domain. Our approach performs parameter efficient multi-domain adaptation without  requiring multiple one-to-one domain adaptation networks.
\\
Cross-domain adaptation has been successfully used for one-to-one domain adaptation and medical image segmentation~\cite{chartsias2017adversarial,huo2018synseg,jiang2018tumor}. These methods produce image-to-image (I2I) translations using the generative adversarial networks (GAN)~\cite{goodfellow2014generative}, including the cycle GAN~\cite{zhu2017unpaired}. GANs map random noise into output images by modeling the intensity distribution of a target domain. But GANs make two unrealistic assumptions for model convergence; the discriminators have infinite capacity to drive the generators, and very large number of training samples are available~\cite{aroraICML17}. The work in~\cite{qi2019loss} showed that the use of appropriate priors like Lipschitz densities are needed to constrain GAN training and avoid mode collapse. 
\\
Bijection constraints, which regularize GAN training by matching a joint distribution of an image and a latent distribution~\cite{zhuNIPS17Bicycle,DonahueKD16}, have shown  feasibility to model diverse intensity and color variations within the same modality. However, these constraints do not specify the dependency (or correlation) structure between the image and the latent distribution and are not guaranteed to cover the different modes in the required target task~\cite{PengICML19,LiuFeatDisNIPS2018}. Prior methods that combined image and feature-level constraints as in ~\cite{hoffman2017,jiang2018tumor,zhu2017unpaired,huo2018synseg} are intrinsically one-to-one mappers, and extension to multi-domain mapping would require $\mathbb{O}(n\times(n-1))$ networks, which is not computationally feasible for practical applications. 
\\
Disentangling methods~\cite{huang2018multimodal,LiuCVPR18}, which extract a domain-invariant content and domain specific style have shown generalizable classification performance on target domains while reducing mode collapse. Example approaches include, task-relevant losses~\cite{PengICML19}, domain classifier losses~\cite{lee2018diverse,choi2018stargan,LiuCVPRMultiTask2018}, and domain generalization methods that match style features with a known prior assumed to span across a number of seen and unseen domains~\cite{LiuFeatDisNIPS2018,PengICML19}. However, in addition to requiring domain-specific style encoders~\cite{lee2018diverse,choi2018stargan}, these methods use image-level matching losses, which is insufficient to model translation of multiple organs that transform differently with respect to one another across the imaging domains. 
\\
\textbf{Key improvements and differences: \/}\rm We used one universal content encoder and one variational auto-encoder to extract image content and style code from multiple domains. The style code is converted into a vector of latent style scales that modulate the generator filters processing content features; target domain code is injected into the generator for synthesizing target modality. Our approach requires no additional networks for encoding domain-specific style features, and requires similar number of networks and parameters as a one-to-one mapper. To our best knowledge, ours is the first approach to perform such scalable multi cross-modality adaptation for disparate medical imaging modalities.  %\textcolor{red}{The style code is used to learn style scale feature to scale the generator feature via a latent scale module (LS). 
%Domain code is injected into both generator and discriminator via a bias center instance normalization (BCIN)\cite{yu2018multi} layer for multiple domain translation training.} 
We improve upon mode-seeking constraints~\cite{maoCVPR2019} to reduce mode collapse by introducing a joint distribution discriminator that combines images with their generated segmentation probability maps to compute domain mismatches. This preserves organs geometry and appearance.\\ \textbf{Contributions:\/}\rm
\begin{enumerate}[topsep=\parskip]
    \item A compact cross-modality  feature disentangling approach for diverse medical imaging modalities adaptation, 
    \item An end-to-end multi-domain translation and unsupervised cross-modality segmentation network, and 
    \item A new joint distribution (image, segmentation map) discriminator to force preservation of multiple organ appearance on generated images. % in multiple domains for better preserving intensity statistics of multiple organs and unsupervised cross-modality multi-organ segmentation. 
    %\item We present extensive evaluation and comparisons of both translation and multi-organ segmentation against multiple methods. %an approach to produce a continuous sampling of the generated images as a combination of the domain codes.
\end{enumerate}
\begin{figure}[t]
	\centering
	\includegraphics[width=\textwidth]{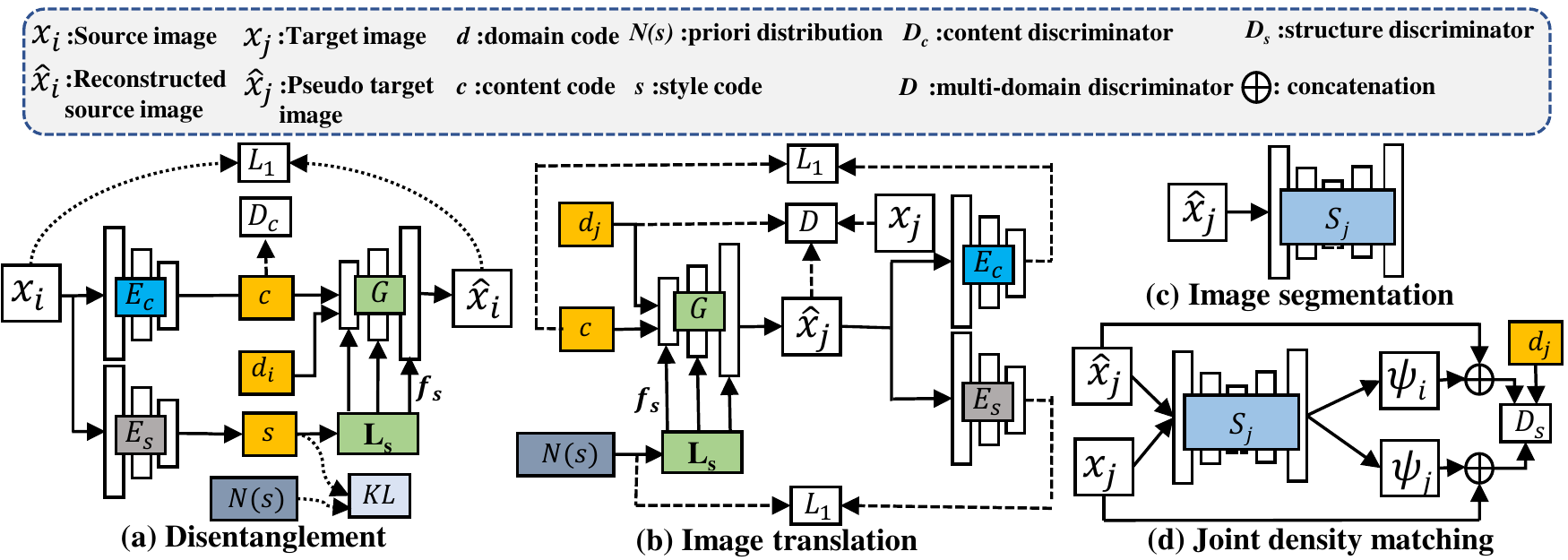}
	\caption{\small{\textcolor{black}{Many-to-many domain translation and multi-domain MRI segmentation with joint distribution discriminator.}}} \label{fig1:method_overview}
\end{figure}
%\textbf{Contributions:\/}\rm

\section{Method}
\subsection{Notation: \/}\rm
Bold letters denote a matrix, \textcolor{black}{$\mathbf{x}, \mathbf{X}$}; mapping functions are denoted by non-bold letters, e.g., $E_c : \mathbf{x_i} \rightarrow \mathbf{c}$, vectors are indicated by italicized letters, e.g., \textit{d\/}\rm. $\mathbf{X}=\cup_{k=1}^{N} \mathbf{X}_{k} \in \mathbb{R}^{H \times W}$ are the set of $N$ domains, $H$ is the height and $W$ is the width of the images. The domain code is represented using one-hot-coding, and is denoted by $\mathbf{d}=\cup_{k=1}^{N}\mathbf{d}_{k}$. Source domain is indicated by $i$, the target domain by $j$, and the transformed image by $\hat{x}_{j}$. Domain-invariant content encoder is denoted by $E_{c}: x_{i} \rightarrow \mathbf{c}$, where \textcolor{black}{$c \in \mathbb{R}^{H^{'}\times W{'} \times C}$}, and $\{H^{'}, W^{'},C\}$ are the height, width of the convolved feature image and number of content features, respectively. The style encoder is denoted by $E_{s}: x_{i} \rightarrow s$, where $s \in \mathbb{R}^{1\times C_s}$, $C_s$ is the number of style features. $G$ is the generator. \textcolor{black}{$L_s$ is the latent scale layer that maps style code s into a vector of latent style scales $f_s \in \mathbb{R}^{1\times F}$, where $F$ is the number of filters in the generator.} $D_{c}$ and $D$ are content and \textcolor{black}{multi-modality} GAN discriminators. $D_{s}$ is the joint distribution (image, segmentation probability map) discriminator. $S_i$ is the segmentator for modality $i$.
\subsection{Feature disentanglement and image translation}
\subsubsection{Style and content feature disentanglement} Disentangled image content and style features (Fig.~\ref{fig1:method_overview}a) are computed using a sequence of convolutional layers and a variational auto-encoder (VAE)~\cite{kingma2013auto}, respectively. Assuming that \textcolor{black}{a latent Gaussian prior} $(z \in \mathcal{N}(0, \textcolor{black}{\textit{I}}))$ spans the styles of all the domains, the encoder $E_s$ extracts a style code that matches with this prior using KL-divergence. The style code is then transformed into latent style scale $f_s$ by a latent scale (LS) layer~\cite{alharbi2019latent}. The latent style scale for each domain is learned. It modulates (as a multiplier) the strength of the various generator filters processing the image content features to produce the desired I2I translation. The vector of latent style scale outputs are shown as multiple outputs for LS in Fig~\ref{fig1:method_overview}(a). These are combined with generator filters using residual blocks. The domain code is injected through channel-wise concatenation with the content code. 
A self-reconstruction loss is combined with the KL-divergence loss in the VAE as:
\begin{equation}
\setlength{\abovedisplayskip}{1pt}
\setlength{\belowdisplayskip}{1pt} 
\textcolor{black}{L_{VAE} = \underset{i}\sum\underset{x_i \sim X_i}E[D_{KL}(E_s(x_i)||z)] + \|\hat{x_i} - x_i \|_{1}.}
\end{equation}  
\textcolor{black}{where $\hat{x_i}$=$G(E_c(x_i),L_s(E_s(x_i)),d_i)$.} The domain invariant content features are produced using an adversarial training by:
\begin{equation}
\setlength{\abovedisplayskip}{1pt}
\setlength{\belowdisplayskip}{1pt} 
  L_{adv}^{c} = \underset{i}\sum\underset{j;j\neq i}\sum\underset{x_i \sim X_i\atop x_j \sim X_j}\E[log(D_c(E_c(x_i)))]+\E[1-log(D_c(E_c(x_j)))],
\end{equation}
%\bumpup
\textbf{Image translation losses:} We compute content reconstruction, latent code regression, \textcolor{black}{domain adversarial loss} and mode-seeking losses to constrain multi-domain I2I translation (Fig.~\ref{fig1:method_overview} b).\\ 
\textbf{Content reconstruction loss} forces content preservation in the transformation $\hat{x}_{j} =  G(E_{c}(x_i), L_s(E_{s}(x_j)), d_j)$ upon transforming an image $x_i$ from domain $i$ to $j$. This is computed as:
\begin{equation}
\setlength{\abovedisplayskip}{1pt}
\setlength{\belowdisplayskip}{1pt} 
\begin{split}
L_{c} = \underset{i}\sum\underset{j;j\neq i}\sum\underset{x_i \sim X_i\atop x_j \sim X_j} \E \| E_{c}(x_{i}) - E_{c}(\hat{x}_{j})\|_{1}. 
%\E_{x \sim S, y \sim T}[||E_s(G(E_c(x),E_s(y),d_y)-E_s(y)||_1)]+ \\
%& \E_{x \sim S, y \sim T}[||E_s(G(E_c(x),E_s(y),d_y)-E_c(x)||_1)]
\end{split}
\end{equation}
\textbf{Latent code regression loss \/} constrains the generator to produce unique mappings of image for a given latent code $z$. This is accomplished through a reverse mapping that produces point-wise estimates of latent code as done in ~\cite{zhuNIPS17Bicycle,lee2019drit++}. The latent regression error is then computed as:
\begin{equation}
\setlength{\abovedisplayskip}{1pt}
\setlength{\belowdisplayskip}{1pt} 
\begin{split}
%L_{lr} = \underset{x \sim X_i} \E \| z - E_{s}(G(E_{c}(x), E_{s}(x), d_i))\|_{1}. \\
\textcolor{black}{L_{lr} = \underset{i}\sum\underset{j;j\neq i}\sum\underset{x \sim X_i} \E \| z - E_{s}(G(E_{c}(x), L_s(z), \textcolor{black}{d_j}))\|_{1}.} 
%\E_{x \sim S, y \sim T}[||E_s(G(E_c(x),E_s(y),d_y)-E_s(y)||_1)]+ \\
%& \E_{x \sim S, y \sim T}[||E_s(G(E_c(x),E_s(y),d_y)-E_c(x)||_1)]
\end{split}
\end{equation}
\textbf{Domain adversarial loss \/}\rm
 is computed to distinguish the generated images from the distribution of the individual domains \textcolor{black}{using styles coded from $x_j$ and sampled from $\mathcal{N}(0, \textit{I})$}. This is computed as:
%\begin{equation}
%\setlength{\abovedisplayskip}{1pt}
%\setlength{\belowdisplayskip}{1pt} 
%\begin{split}
%\underset{G}{\mathrm{min}}\ \underset{D}{\mathrm{max}}\  L_{adv} = \underset{i,j; i\neq j} \sum \underset{y \sim X_{j}} \E[log (D_{j}(y))] + \underset{x \sim X_{i}} \E [log(1 - D_{j}(G(x)))].\\
%\end{split}
%\end{equation}
\begin{equation}
\setlength{\abovedisplayskip}{1pt}
\setlength{\belowdisplayskip}{1pt} 
\begin{split}
\underset{G}{\mathrm{min}}\ \underset{D}{\mathrm{max}}\  L_{GAN} = &\underset{i} \sum \underset{j; j\neq i} \sum \{\underset{x_j \sim X_{j}} \E[log (D(x_j,d_j))] +\\  & \underset{x_i \sim X_{i}} \E [\textcolor{black}{\frac{1}{2}}log(1 - D(G(E_c(x_i),L_s(E_s(x_j)),d_j),d_j))] + \\
& \underset{z \sim N(0,1)} \E [\textcolor{black}{\frac{1}{2}}log(1 - D(G(E_c(x_i),L_s(z),d_j),d_j))\}
\end{split}
\end{equation}
Multi-domain translation is additionally stabilized by employing bias center instance normalization (BCIN) in both the generator and discriminators, as it has been shown to improve the consistency and diversity in the generated I2I transformations\cite{yu2018multi}. BCIN also allows the domain labels $d_j$ $\in$ $\mathbf{d}$ to be directly injected into G and D, which eliminates additional computations otherwise needed for calculating \textcolor{black}{domain} classification loss via the discriminator as done in \cite{choi2018stargan,lee2019drit++}. \\
\textbf{Mode seeking loss: }As proposed in~\cite{maoCVPR2019}, we compute a mode-seeking loss to prevent mode collapse such that the chances of producing the same image from two different latent vectors is reduced. Briefly, given \textcolor{black}{two random latent style codes $z_{1}, z_{2} \in \mathcal{N}(0,I)$}, an image $x_i \sim X_{i}$, target domain code $d_j$ and a generator $G$, mode seeking regularization maximizes the ratio of the distance between the generated images and the corresponding latent vectors as: 
\begin{equation}
\setlength{\abovedisplayskip}{1pt}
\setlength{\belowdisplayskip}{1pt} 
%L_{ms} = \max_{G} \bigg(\underset{i}\sum\underset{j;j\neq i}\sum\frac{\mathb{d}_\mathb{I}(G(E_{c}(x_i),L_{s}(z_{1}),d_{j}),G(E_{c}(x_i),L_{s}(z_{2}),d_{j}))}{d_{z}(z_{1},z_{2})}\bigg)
L_{ms} = \max_{G} \bigg(\underset{i}\sum\underset{j;j\neq i}\sum\frac{d_{I}(G(E_{c}(x_i),L_{s}(z_{1}),d_{j}),G(E_{c}(x_i),L_{s}(z_{2}),d_{j}))}{d_{z}(z_{1},z_{2})}\bigg)
\end{equation}
Finally, a new joint distribution discriminator is employed as described in the next Subsection to constrain the multi-domain translation for segmentation.  
%\bumpup
\subsection{Segmentation}
%\bumpup
CT is the source domain containing expert segmentations, $\{X_i, Y_i\}$, and the target MRI domain only contains the image sequences for training $\mathbf{X}=\cup_{j=1,j\neq i}^{N} \mathbf{X}_{j}$. 
Separate multi-organ segmentation networks $S_j$ are trained for each target modality $j$ (Fig.~\ref{fig1:method_overview}(c)) by using transformed images $\hat{x}_j$=$G(E_c(x_i),L_s(z),d_j)$, \textcolor{black}{obtained via randomly} sampled style z during segmentation training. Cross entropy loss was used to optimize these networks as:
\begin{equation}
    \begin{split}
    \setlength{\abovedisplayskip}{1pt}
    \setlength{\belowdisplayskip}{1pt}
    L_{seg} = \underset{j,j\neq i}{\sum} \underset{\hat{x}_j \sim \hat{X}_j, l_i \sim Y_i} \E [logP(l_i|S_{j}(\hat{x}_{j}))].  \label{eqn:segmentation_loss}
    \end{split}
\end{equation}
\begin{comment}
\begin{equation}
    \begin{split}
    \setlength{\abovedisplayskip}{1pt}
    \setlength{\belowdisplayskip}{1pt}
    L_{seg} = \underset{j,j\neq ic}{\sum} \underset{\hat{x}_j \sim X_{ic}^{j}, l_c \sim Y_c} \E [logP(l_c|S_{j}(\hat{x}))],  \label{eqn:segmentation_loss}
    \end{split}
\end{equation}
\end{comment}
\textbf{Joint distribution structure discriminator}
Prior work~\cite{jiang2018tumor} has shown that modality hallucination occurs while transforming highly disparate modalities when GAN training is not regularized to model the structures of interest. Therefore, we introduce a new adversarially trained joint distribution structure discriminator $D_s$ (Fig.~\ref{fig1:method_overview}(d)) that explicitly conditions image generation by focusing domain mismatch detection within the structures of interest. This is done by treating images and their segmentation probabilities as a joint distribution, and implemented as channel-wise concatenation. Voxel-wise segmentation probability maps (channel-wise accumulation except the 0-th channel that corresponds to background label) $\hat{\psi}_{j}$ for the CT to MRI translated and the unrelated real MRIs $\psi_{j}$ are obtained from the \textit{SoftMax\/} \rm operation of the segmentation network $S_j$. $D_s$ uses the domain code $d_j$ to compute domain mismatch as:
\begin{equation}
\begin{split}
\textcolor{black}{
L_{st}= \underset{j;j\neq i}{\sum}\{\E_{x_j \sim X_j} [log(D_{s}(x_j, \psi_{j},d_j))]  +  
 \E_{\hat{x}_j \approx X_{j}} [log(1-D_{s}(\hat{x}_j,\hat{\psi}_{j},d_j))]}\}
\end{split} 
\label{eqn:Joint_D_psi}
\end{equation}  
The total loss is computed as:
\begin{equation}
\setlength{\abovedisplayskip}{1pt}
\setlength{\belowdisplayskip}{1pt}
L_{total}=L_{GAN}+{\lambda_{vae}L_{VAE}}+\lambda_{c}{L_{adv}^{c}}+\lambda_{lr}{L_{lr}}+\lambda_{ms}{L_{ms}}+\lambda_{st}{L_{st}}+\lambda_{seg}{L_{seg}}  
\label{eqn:Total loss}
\end{equation}
%\bumpup
%\bumpup
%\bumpup
\subsection{Implementation details and network structure}
%\bumpup
All networks were implemented using the Pytorch library and were trained on Nvidia GTX V100 with 16 GB memory. The ADAM algorithm\cite{kingma2014adam} with an initial learning rate of 2e-4 and batch size of 1 was used during training. We set $\lambda_{vae}$=1, $\lambda_{c}$=1, $\lambda_{lr}$=10, $\lambda_{ms}$=1, $\lambda_{st}$=1 and $\lambda_{seg}$=5 in the training. The learning rate was kept constant for the first 50 epochs and decayed to zero in the next 50 epochs. In order to ensure stable training, the encoders, G and VAE networks are trained cooperatively and optimized at a different iteration than the discriminators and segmentors. 
\\
The content encoder $E_c$ is a fully convolutional network for projecting the images into a spatial feature map. The feature map retains the spatial structure by using a small output stride of 2. The style encoder $E_s$ is composed of several convolution and pooling layers followed by global pooing and fully connected layers, with the output layer implemented using a reparameterization trick as done in~\cite{kingma2013auto}. The latent scale module $L_s$ consists of 5 fully connected layers with \textit{tanh\/} \rm operation to produce style scale features within a range of [-1, 1]. The generator $G$ is composed of 5 residual blocks with BCIN used for domain code injection. The segmentation networks $S$ are implemented using a standard Unet\cite{ronneberger2015u}. Content discriminator $D_c$ is implemented using PatchGAN. The joint density structure discriminator $D_s$ and multi-domain discriminator $D$ are implemented using PatchGAN with BCIN  used for domain injection. Details of all networks are included in the supplementary documents. 
\section{Experiments and Results}
\textbf{Dataset:} We used 20 MRIs (T1w and T2w MRI) from the Combined Healthy Abdominal Organ Segmentation (CHAOS) challenge data~\cite{CHAOS2019} and a completely different set of 30 patients with expert-segmented CT scans from~\cite{landman2015miccai} for the analysis. CHAOS CT scans only have liver segmentations. Ten MRIs were used in training (without expert segmentations) and validation (with expert segmentations) while the remaining 10 MRIs were held out for independent testing.\\
Networks were trained using 256$\times$256 pixels image patches obtained from 14038 individual CT slices, and from 8000 T1w and 7872 T2w MRI slices, respectively. All MRIs were acquired from 1.5T Philips scanners, with a resolution of 256$\times$256, pixel spacing [1.36mm to 1.89mm] and slice thickness [5.5mm to 9mm], and processed with bias field correction. CTs had a resolution of 512$\times$512 pixels, a pixel resolution [0.7 to 0.8mm], and a slice thickness [3mm to 3.2mm]. CT and MR image sets were normalized in range  -1 to +1 prior to training and testing.\\
%\bumpdown
\textbf{Experiments: }
Performance comparisons were done against state-of-the-art multi-domain disentanglement based translation methods, DRIFT++~\cite{lee2019drit++} and  starGAN~\cite{choi2018stargan}. Translation accuracy was evaluated using cyclic  reconstruction \textcolor{black}{\cite{qi2019loss}} of CT going through T1w and T2w MRI using mean absolute error (MAE) and through T-SNE cluster distances between translated and real MRIs. \\
Segmentation performance was compared against multiple translation based segmentation methods including Cycada~\cite{hoffman2017}, CycleGAN~\cite{zhu2017unpaired}, variational auto-encoder based method UNIT~\cite{hou2017unsupervised}, synergistic feature encoder (SIFA)~\cite{chen2019synergistic}, and the    SynSeg~\cite{huo2018synseg} methods. Dice similarity coefficient (DSC) was used to measure accuracy. All methods were trained from scratch with identical image sets and subject to reasonable hyperparameter optimization for equitable comparisons.
\begin{figure}[t]
\centering
\includegraphics[width=0.9\textwidth]{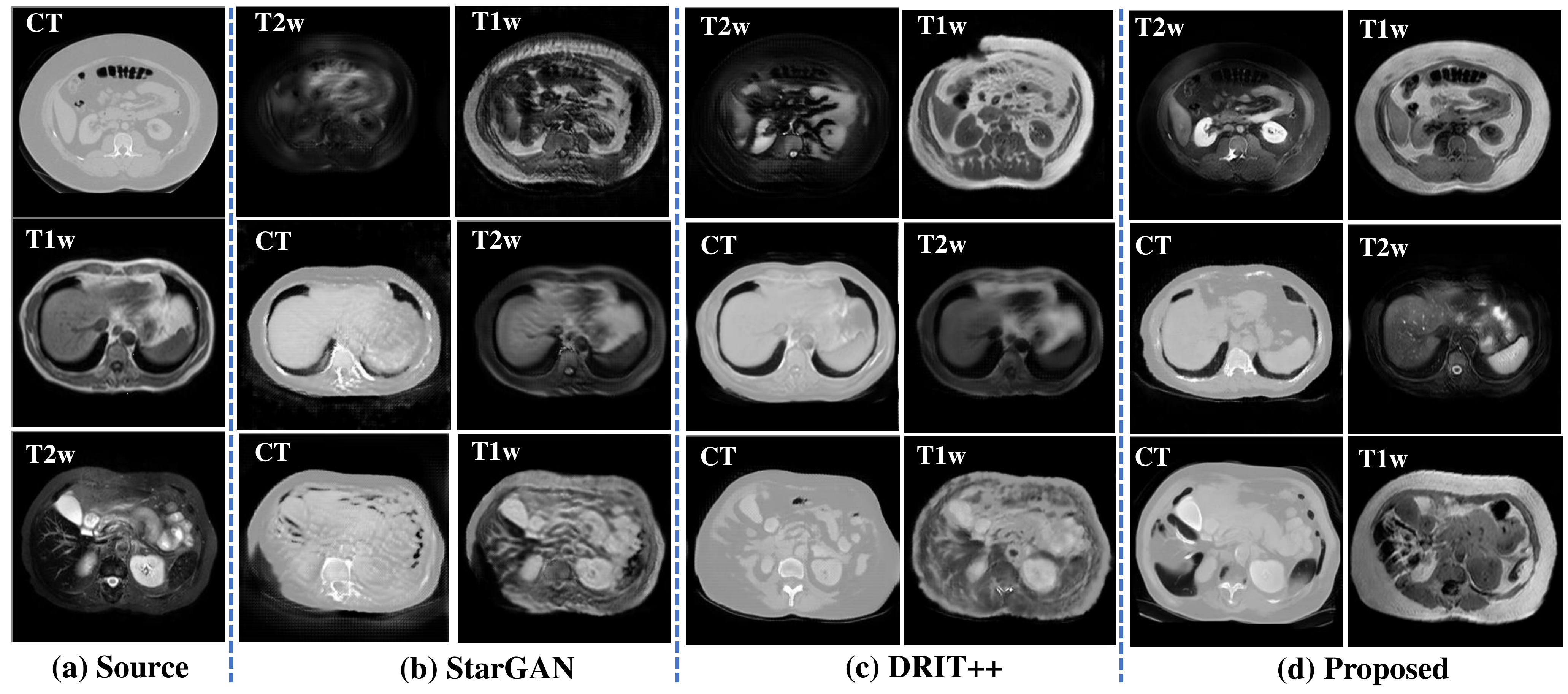}
\caption{\small{Multi-domain many-to-many translation. }} 
\label{fig1:translation_result_M2M}
\end{figure}
%\bumpup
%CT->T1->T2-CT
%\bumpup
%\bumpup
\subsection{Results and discussion}
%\bumpup
\subsubsection{Image translation} 
\begin{figure}[t]
\centering
\includegraphics[width=0.95\textwidth]{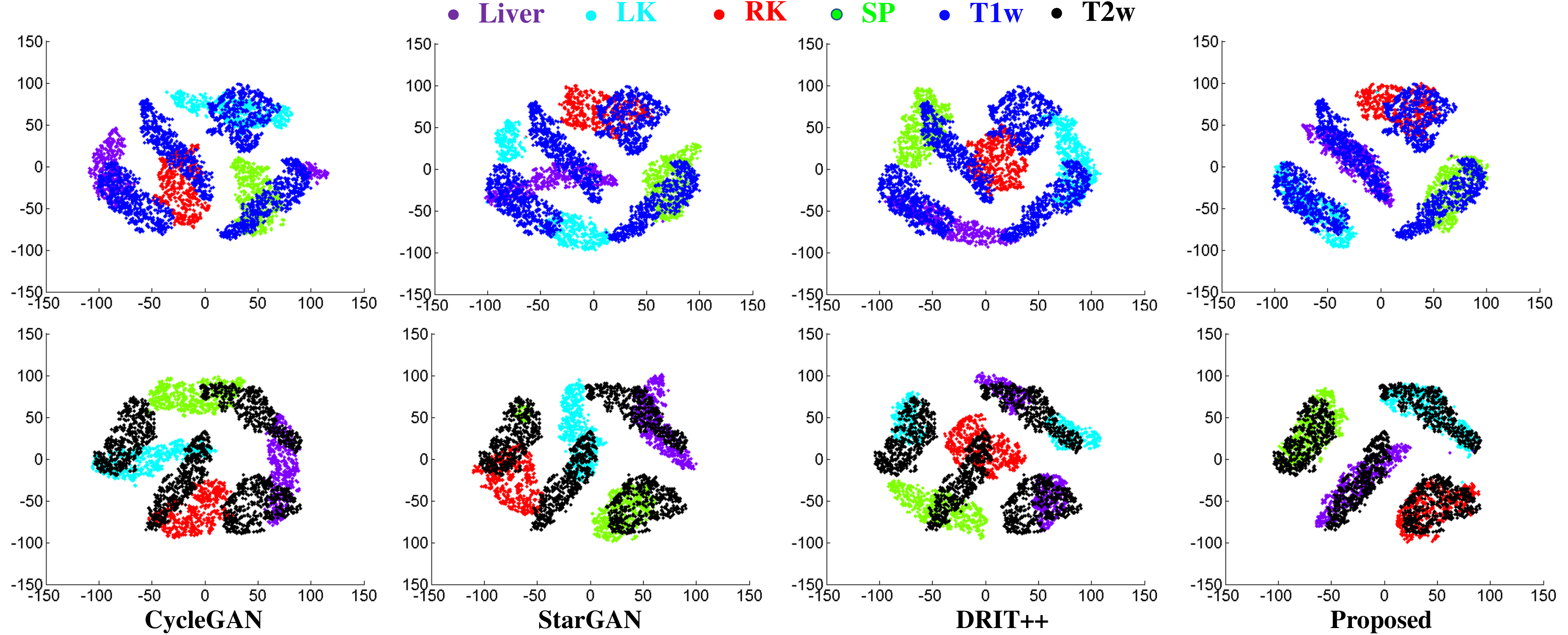}
\caption{\small{TSNE clusters computed from the generated and real T1w, T2w MRI of organs.}} 
\label{fig1:TSNE}
\end{figure}

\begin{figure}[t]
\centering
\includegraphics[width=1\textwidth]{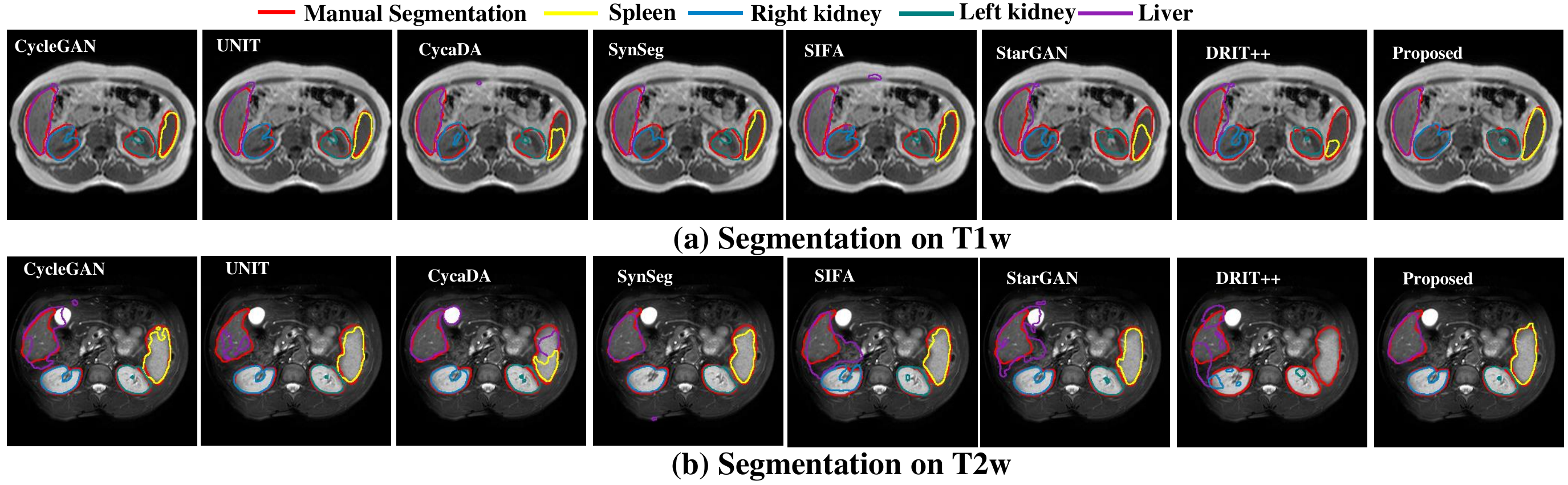}
\caption{\small{Segmentation results comparing the various methods to the proposed method.}}
%Expert delineations are shown in red while the algorithms' delineations for the individual organs are shown in different colors.}} 
\label{fig1:seg_overlay}
\end{figure}

Fig.~\ref{fig1:translation_result_M2M} shows example  translation from all three modalities using our and other methods. As shown, both StarGAN\cite{choi2018stargan} and DRIT++\cite{lee2019drit++} produce less accurate translation for all three modalities compared with our method. Fig.~\ref{fig1:TSNE} shows T-SNE clusterings computed from the signal intensities of the various organs of interest using multiple methods and overlayed with the distribution of real T1w and T2w MRIs. The average cluster distance to the corresponding real MRI using our method was 5.05 and 14.00 for T2w and T1w MRIs, respectively. The average cluster distances of comparison methods were much higher with CycleGAN\cite{zhu2017unpaired} of 73.90 for T2w, 101.37 for T1w, StarGAN\cite{choi2018stargan} of 73.39 for T2w and 77.49 for T1w, and DRIT++\cite{lee2019drit++} of 87.32 for T2w and 70.73 for T1w MRIs, respectively. Furthermore, as shown in Table~\ref{tab:RE_error}, our method produced the lowest reconstruction error for one step and two-step reconstructions.\\
\textbf{Segmentation } Table~\ref{tab:Seg_result} shows the segmentation accuracies achieved by our and multiple methods.  We also evaluated the performance when a network trained on CT was directly applied to T1w MRI and T2w MRIs. As seen, the performance worsens when using the network trained on CT dataset alone. We evaluated the performance of the approach without the joint density structure discriminator in order to evaluate its use for segmentation performance. As shown, adding the joint density discriminator improved the segmentation performance. Overall, our approach produced an average accuracy of 0.85 for T1w and 0.90 for T2w MRI, which was higher than all other compared methods and only slightly lower than fully supervised training with T1w MRI, and slightly improved over supervised training for T2w MRI.  Fig.~\ref{fig1:seg_overlay} shows example segmentations produced for T1w and T2w images using all the compared methods against expert segmentation.\\
\textbf{Ablation tests } We evaluated the impact of  mode-seeking loss~\cite{maoCVPR2019} and joint distribution matching losses introduced in this work for T2w MRI segmentation. The two step reconstruction error increased 27\% and 19\%, while segmentation accuracy dropped by 8\% and 7\% when removing mode seeking loss and structure discriminator loss, respectively.\\
\textbf{Discussion } We developed a multi cross-modality adaptation-based unsupervised segmentation approach that uses only single encoder/decoder for producing one to many mapping. Prior feature disentanglement medical image analysis methods used separate one-to-one mapping based segmentors\cite{yang2019unsupervised,chen2019synergistic,chartsias2018factorised}, wherein each additional modality would require one additional encoder and decoder. Also, multi-domain adaptation was often done to handle scanner-related imaging variation in that same modality \cite{yang2019unsupervised,ouyangMICCAI2019,chartsias2018factorised}. Our work also extends unsupervised image-level classification~\cite{huang2018multimodal,LiCVPR2018}
to unsupervised multi-domain,  
multiple organ segmentation. Our results showed improved segmentation and reconstruction performance against the compared methods. 
%\bumpup
%\bumpup
%\bumpup
%\bumpup
\begin{table*}[t]
\centering{\caption{\small{Reconstruction Error.}} 
	\label{tab:RE_error} 
	\centering
	%\scriptsize
	%\tiny
	\scriptsize
	%\footnotesize
	%\small
	%\normalsize
	%\large
	%\Large
	%\LARGE
	%\huge
	%\Huge
	\centering
\begin{tabular}{c|c|c|c|c|} 
			\hline
			
			\hline
	\multirow{2}{*}{Method}   &\multicolumn{2}{c}{One-step reconstruction } & \multicolumn{2}{|c|}{Two-step reconstruction}\\
	\cline{2-5}
	{} & \multicolumn{1}{c|}
	{CT $\rightarrow$T1w$\rightarrow$CT}& {CT$\rightarrow$ T2w$\rightarrow$CT}& {CT$\rightarrow$T1w$\rightarrow$T2w$\rightarrow$CT}& {CT$\rightarrow$T2w$\rightarrow$T1w$\rightarrow$CT} \\
	\hline
\multirow{1}{*}{StarGAN}&   {2.30$\pm$1.37}   &   {2.56$\pm$1.30} &   {5.56$\pm$2.44}  &   {6.55$\pm$2.07} \\
	\hline
\multirow{1}{*}{DRIT++}&   {2.32$\pm$1.00}   &   {2.53$\pm$1.08} &   {6.13$\pm$1.40}  &   {6.04$\pm$2.23} \\
	\hline
\multirow{1}{*}{Proposed}&   {\textbf{1.34$\pm$0.40}}   &   {\textbf{1.20$\pm$0.39}} &   {\textbf{2.36$\pm$0.44}}  &   {\textbf{2.27$\pm$0.67}} \\
	\hline
	\end{tabular}} 
\end{table*}
\begin{table*}[t]
\centering{\caption{\small{Overall segmentation accuracy on CHAOS dataset (In-Phase). Liver-LV, Spleen-SP, Left kidney-LK, Right kidney-RK.}} 
	\label{tab:Seg_result} 
	\centering
	%\scriptsize
	%\tiny
	\scriptsize
	%\footnotesize
	%\small
	%\normalsize
	%\large
	%\Large
	%\LARGE
	%\huge
	%\Huge
	\centering
\begin{tabular}{c|c|c|c|c|c|c|c|c|c|c} 
			\hline
			
			\hline
	\multirow{2}{*}{Method}   &\multicolumn{5}{c}{T1w } & \multicolumn{5}{|c}{T2W}\\
	\cline{2-11}
	{} & \multicolumn{1}{c|}
	{\textcolor{white}{aa}LV\textcolor{white}{aa}}& {\textcolor{white}{aa}SP\textcolor{white}{aa}}& {\textcolor{white}{aa}LK\textcolor{white}{aa}}& {\textcolor{white}{aa}RK\textcolor{white}{aa}}& {Avg} & {\textcolor{white}{aa}LV\textcolor{white}{aa}}& {\textcolor{white}{aa}Sp\textcolor{white}{aa}}& {\textcolor{white}{aa}LK\textcolor{white}{aa}}& {\textcolor{white}{aa}RK\textcolor{white}{aa}}& {Avg} \\
	\cline{3-11}	 
	\hline
\multirow{1}{*}{MRI supervised}&   {0.91}   &   {0.86} &   {0.82}  &   {0.83} &{0.86}  &  {0.92}   &   {0.87} &   {0.91}  &   {0.90}&   {0.90}\\
\hline
\multirow{1}{*}{CT only }&{0.00} &   {0.00}   &   {0.00} &   {0.00}  &   {0.00}     &   {0.01} &   {0.08}  &   {0.29}&   {0.22}&  {0.15}\\
\hline
\hline
	\multirow{1}{*}{CycleGAN\cite{zhu2017unpaired}}&   {0.82}   &   {0.83} &   {0.63}  &   {0.61} &{0.72}  &  {0.86}   &   {0.75} &   {0.88}  &   {0.87}&   {0.84}\\
	\hline
	\multirow{1}{*}{UNIT\cite{hou2017unsupervised}}&   {0.89}   &   {0.81} &   {0.64}  &   {0.62} &{0.74}  &  {0.87}   &   {0.76} &   {0.91}  &   {0.88}&   {0.86}\\
	
	\hline
	\multirow{1}{*}{CycaDA\cite{hoffman2017}}&   {0.85}   &   {0.73} &   {0.71}  &   {0.70}  &{0.75} &  {0.88}   &   {0.68} &   {0.86}  &   {0.86}&   {0.82}\\
	\hline	
	\multirow{1}{*}{SynSeg\cite{huo2018synseg}}&   {0.89}   &   {\textbf{0.85}} &   {0.73}  &   {0.70} &{0.79}  &  {0.88}   &   {0.77} &   {0.89}  &   {0.85}&   {0.85}\\
	\hline	
	\multirow{1}{*}{SIFA\cite{chen2019synergistic}}&   {\textbf{0.90}}   &   {\textbf{0.85}} &   {0.77}  &   {0.78} &{0.83}  &  {0.89}   &   {0.77} &   {0.90}  &   {0.89}&   {0.86}\\
	\hline	
    \hline
	\multirow{1}{*}{StarGAN\cite{choi2018stargan}}&   {0.76}   &   {0.60} &   {0.56}  &   {0.67}  &{0.65} &  {0.83}   &   {0.69} &   {0.69}  &   {0.73}&   {0.74}\\ 
	\hline
	\multirow{1}{*}{DRIT++\cite{lee2019drit++}}&  {0.83}& {0.63}   &   {0.61} &   {0.66}       &{0.68} &  {0.73} & {0.81}   &   {0.81} &   {0.82}&   {0.79}   \\ 
	\hline
	\hline
	\multirow{1}{*}{Proposed - $D_{s}$}&{0.89}&{0.79}&{0.74}&{0.73}&{0.79}
	&{0.87}&{0.86}&{0.91}&{0.89}&{0.88}\\     
	\hline
	\multirow{1}{*}{Proposed + $D_{s}$ }&   {\textbf{0.90}}   &   {0.84} &   {\textbf{0.81}}  &   {\textbf{0.83}} &{\textbf{0.85}}  &  {\textbf{0.90}}   &   {\textbf{0.89}} &   {\textbf{0.92}}  &   {\textbf{0.90}}&   {\textbf{0.90}}\\     
	\hline		
	
	\hline
	\end{tabular}} 
\end{table*}	
\section{Conclusion}
%\bumpup
We developed a multi-domain adversarial translation and segmentation method applied to unsupervised multiple MRI sequences segmentation. We showed that a universal multi-domain disentanglement using content and style extractors can produce reasonably accurate multi-domain translation and reasonably accurate segmentation of multiple organs. 
\paragraph{\bf{Acknowledgement.}} This work was supported by the MSK Cancer Center support grant/core grant
P30 CA008748.
{\tiny
	\bibliographystyle{splncs}
}
%\bumpup
\bibliography{mybibliography}

%Author, F.: Article title. Journal \textbf{2}(5), 99--110 (2016)

%Author, F., Author, S.: Title of a proceedings paper. In: Editor,
%F., Editor, S. (eds.) CONFERENCE 2016, LNCS, vol. 9999, pp. 1--13.
%Springer, Heidelberg (2016). \doi{10.10007/1234567890}

%Author, F., Author, S., Author, T.: Book title. 2nd edn. Publisher,

%Author, A.-B.: Contribution title. In: 9th International Proceedings
%on Proceedings, pp. 1--2. Publisher, Location (2010)

%LNCS Homepage, \url{http://www.springer.com/lncs}. Last accessed 4
%\end{thebibliography}
\end{document}